\documentclass[11pt]{article}
\usepackage{amsmath}
\usepackage{mathrsfs,epsfig,morefloats}
\usepackage{amsfonts}
\usepackage{mathtools}
\usepackage{authblk}
\usepackage{xcolor}
\usepackage{graphicx}
\usepackage{float}
\usepackage{amsthm}
\usepackage{amssymb}
\usepackage{lscape}
\usepackage{rotating}
\usepackage{subfigure}
\usepackage{graphicx}

\usepackage{tabularx}
\usepackage{multirow}

\usepackage{algorithm}

\allowdisplaybreaks

\newenvironment{titlemize}[1]{
  \paragraph{#1}
  \begin{enumerate}}
  {\end{enumerate}}

\usepackage{algpseudocode}
\newcommand\Algphase[1]{
\vspace*{-.7\baselineskip}\Statex\hspace*{\dimexpr-\algorithmicindent-2pt\relax}\rule{\textwidth}{0.4pt}
\Statex\hspace*{-\algorithmicindent}\textbf{#1}%
\vspace*{-.7\baselineskip}\Statex\hspace*{\dimexpr-\algorithmicindent-2pt\relax}\rule{\textwidth}{0.4pt}
}

\usepackage{caption}
\usepackage{lipsum}

\DeclareCaptionFormat{algor}{%
  \hrulefill\par\offinterlineskip\vskip1pt%
    \textbf{#1#2}#3\offinterlineskip\hrulefill}
\DeclareCaptionStyle{algori}{singlelinecheck=off,format=algor,labelsep=space}
\usepackage{hyperref}
\hypersetup{
    colorlinks=true,
    linkcolor=blue,
    filecolor=blue,      
    urlcolor=blue,
citecolor=blue,
}
 \usepackage[sort,comma,authoryear,round]{natbib}

\usepackage[section]{placeins}
\newcommand{\Date}[1]{\def\@Date{#1}}
\def\today{\number\day~\ifcase\month\or
 January\or February\or March\or April\or May\or June\or
 July\or August\or September\or October\or November\or December\fi~\number\year}

\def\be{\begin{equation}}
\def\ee{\end{equation}}
\def\bea{\begin{eqnarray}}
\def\eea{\end{eqnarray}}
\def\bd{\begin{displaymath}}
\def\ed{\end{displaymath}}
\def\bda{\begin{eqnarray*}}
\def\eda{\end{eqnarray*}}
\def\bsm{\begin{small}}
\def\esm{\end{small}}

\def\t0{\theta_0}

\def\ha1{\hat \beta_1}

\def\bnt{\begin{enumerate}}
\def\ent{\end{enumerate}}

\def\bsc{\begin{scriptsize}}
\def\esc{\end{scriptsize}}

\theoremstyle{definition}

\newtheorem{remark}{Remark}

\makeatletter
\newcommand{\figcaption}{\def\@captype{figure}\caption}
\newcommand{\tabcaption}{\def\@captype{table}\caption}
\makeatother

\begin{document}

\title{\bf IFAA: Robust association identification and {\bf I}nference {\bf F}or {\bf A}bsolute {\bf A}bundance in microbiome analyses}

\author{\normalsize Zhigang Li$^a$\thanks{Corresponding author: zhigang.li@ufl.edu.}, Lu Tian$^b$, A. James O'Malley$^c$, Margaret R. Karagas$^d$, Anne G. Hoen$^d$, Brock C. Christensen$^d$, Juliette C. Madan$^d$, Quran Wu$^a$, Raad Z. Gharaibeh$^e$, Christian Jobin$^e$, and Hongzhe Li$^f$\\
\normalsize {$^a$Department of Biostatistics, University of Florida, Gainesville, FL; $^b$Department of Biomedical Data Science, Stanford University, Palo Alto, CA;  $^c$The Dartmouth Institute, Geisel School of Medicine at Dartmouth, Hanover, NH; $^d$Department of Epidemiology, Geisel School of Medicine at Dartmouth, Hanover, NH;  $^e$Department of Medicine, University of Florida,  Gainesville, FL; $^f$Department of Biostatistics, Epidemiology \& Informatics, University of Pennsylvania, Philadelphia, PA.}}

\date{}
\maketitle

\begin{abstract}
The target of inference in microbiome analyses is usually relative abundance (RA) because RA in a sample (e.g., stool) can be considered as an approximation of RA in an entire ecosystem (e.g., gut). However, inference on RA suffers from the fact that RA are calculated by dividing absolute abundances (AA) over the common denominator (CD), the summation of all AA (i.e., library size). Because of that, perturbation in one taxon will result in a change in the CD and thus cause false changes in RA of all other taxa, and those false changes could lead to false positive/negative findings. We propose a novel analysis approach (IFAA) to make robust inference on AA of an ecosystem that can circumvent the issues induced by the CD problem and compositional structure of RA. IFAA can also address the confounding effect of library size and handle zero-inflated data structures. IFAA identifies microbial taxa associated with the covariates in Phase one and estimates the association parameters by employing an independent reference taxon in Phase two. Two real data applications are presented and extensive simulations show that IFAA outperforms other established existing approaches by a big margin in the presence of confounding effect of library size. 
\end{abstract}

\noindent {\sl Keywords}: Compositional data; Differential abundance analysis; High dimension; Microbiome regression; Zero-inflated data.

\section{Introduction}\label{s:intro}
The human microbiome consist of trillions of microorganisms including bacteria, archaea, viruses, and fungi living in and on the human body and play important roles in our health (\citealp{Turnbaugh2007}; \citealp{Consortium2012}; \citealp{Lloyd-Price2017}). Microbial dysbiosis has been linked to a variety of diseases including asthma, infection, and allergy in children (\citealp{Chen2007};  \citealp{Madan2012}; \citealp{Hoen2015}), as well as cancer (\citealp{Reikvam2011}; \citealp{Castellarin2012}) and obesity (\citealp{Turnbaugh2006}; \citealp{Trasande2013}). To quantitatively study the assocation of human microbiome with exposure variables and clinical outcomes, sequencing technologies such as 16s ribosomal RNA gene sequencing \citep{Cole2009} and shotgun metagenomic sequencing \citep{Tringe2005} are employed to quantify the microbiome composition of a sample (e.g., stool, saliva), and then numerical measures for the association of interest can be derived with statistical and computational methods \citep{Lee2015}. Because sequencing data is collected from a sample representing a small proportion of the ecosystem (e.g., gut), the raw sequencing count (i.e., absolute abundance) of a microbial taxon in the sample may not serve as a good estimate for its absolute abundance (AA) in the ecosystem \citep{ancom}. The target of inference is very often the relative abundances (RA) which measure the fractions of microbial taxa in the ecosystem that can be approximated by the observed fractions in the sample (\citealp{Lozupone2005}; \citealp{LaRosa2012}; \citealp{Chen2013}; \citealp{Tang2018}).

Making inference on RA is challenging because perturbation in the abundance of one taxon will cause changes in fractions of all taxa due to  change in the common denominator (CD) for calculating all fractions. We will refer to this as the CD problem hereafter. The CD problem is also related to the compositional structure of RA's under which they are negatively correlated since an increase in one RA will necessarily result in a decrease in another one. Those false changes could generate false positive results or mask the true changes which then lead to false negative results. Existing methods have not been able to adequately resolve this issue. 

Another well-known challenge for making inference on RA comes from the zero-inflated structure of the sequencing data which is also a general challenge for analyzing microbiome data. Many existing methods (\citealp{Chen2013}; \citealp{zig}; \citealp{Lin2014};  \citealp{ancom}) require imputing zero-valued sequencing counts with a positive number such as the Pseudocount of 0.5 or another number which could lead to biased estimates of the RA's. Because the log transformation over the interval (0,1) ranges from negative infinity to zero, this bias can be exaggerated to a surprisingly large value on the commonly used natural-log scale. For example, when RA changes from 0.1 to 0.00001 which corresponds to approximately 1-fold decrease in terms of magnitude on the original RA scale, the log-value of RA changes from $-2.30$ to $-11.51$ corresponding to a 5-fold change in terms of magnitude. Imputation of the zero counts could also be problematic when the sequencing depth (i.e., library size) is a confounder of the association of interest \citep{Weiss2017}. Sequencing depth has a strong correlation with the diversity of microbiome community observed in a sample. For instance, the number of detected OTUs and sequencing depth are highly correlated (with an r-square of 0.92) in the Human Microbiome Project (\citealp{Turnbaugh2007}; \citealp{Consortium2012}) as shown in \cite{zig}. So when comparing two groups, it is possible that one group has some taxa with more zero-valued RA due to smaller library sizes, and consequently imputation of the zero reads could create an artificial bias for the group difference when the true difference is null. 

To address the above challenges, we propose a novel approach to draw inference on the AA of the ecosystem instead of the RA. The new approach will avoid the aforementioned CD problem associated with RA and get rid of the issue induced by the compositional structure of RA's. Unlike many existing methods, this new method does not require imputing zero although it can be used to analyze microbiome data after zero values are imputed with a pseudo count or any other number. The new algorithm consists of two phases with Phase 1 to identify the taxa whose AA are associated with the covariates of interest and Phase 2 to estimate the association parameters. Both phases utilize the ratios of non-zero AA observed in the samples. The advantage of using the ratios is that it can remove the impact of sequencing depth in the model because the sequencing depth is cancelled out in calculating the ratios. The key idea of phase 1 is that the ratio of two taxa should be independent of the covariates of interest if the two taxa are both independent of the covariates, and the ratio should be associated with the covariates if any one of the two taxa is associated with the covariates. This will allow for identification of the set of taxa (set A) that are associated with the covariates as well as the other set of taxa (set B) that are not associated with the covariates, and then in phase 2 we are able to quantify the associations between AA and the covariates for all taxa in set A with point estimates and confidence intervals by using a reference taxon that is indepdent of the covariates. Our approach can also remove the confounding effect of sequencing depth because the ratio of two taxa abundances does not depend on the sequencing depth, and thus it can not be a confounder in the model. By incorporating regularization methods, our approach can handle high-dimensional microbiome data as well as high-dimensional covariates data.

We organize this paper as follows. Model and notations are presented in Section \ref{s:model}. Algorithms for identifying sets A and B and for parameter estimation are provided in Section \ref{s:estimation} followed by an extensive simulation study under different scenarios to assess the performance of our approach in comparison with other established existing approaches in Section \ref{s:simu}. We showcase the application of our new approach to two real studies in comparison with existing approaches in Section \ref{s:app} followed by the discussion in Section \ref{s:discu}.

\section{Model and Notation}\label{s:model}
\subsection{Multivariate zero-inflated log-normal distribution}\label{ss:ZILoN}
Suppose there are $N$ subjects and $K+1$ taxa of interest. Let $\mathcal{Y}_i=(\mathcal{Y}^1_i,\mathcal{Y}^2_i,...,\mathcal{Y}^K_i,\mathcal{Y}^{K+1}_i)$ denote the true microbial taxa absolute abundance (ie, counts) in the ecosystem (eg, gut) of the $i$th subject, $i=1,...,N$. The subject index $i$ will be suppressed for simplicity in this section. To describe the microbial abundance distribution, we propose a multivariate zero-inflated log-normal distribution that can account for the zero-inflated structure. It is a two-part distribution with a discrete part and a continuous part. The discrete part provides the probabilities governing the probabilities of taxa abundance being zero or non-zero: 
\begin{align*}
P(\mathcal{Y}^1>0,\mathcal{Y}^2=0,...,\mathcal{Y}^K=0,\mathcal{Y}^{K+1}=0)=p_1 \hspace{0.3cm}\\
P(\mathcal{Y}^1=0,\mathcal{Y}^2>0,...,\mathcal{Y}^K=0,\mathcal{Y}^{K+1}=0)=p_2  \hspace{0.3cm}\\
\vdots \hspace{5cm}\\
P(\mathcal{Y}^1=0,\mathcal{Y}^2=0,...,\mathcal{Y}^K=0,\mathcal{Y}^{K+1}>0)=p_{K+1} \\
\vdots \hspace{5cm}
\end{align*}
\begin{align*}
P(\mathcal{Y}^1=0,...,\mathcal{Y}^{k_1-1}=0,\mathcal{Y}^{k_1}>0,\mathcal{Y}^{k_1+1}=0,...,\mathcal{Y}^{k_L}>0,...,\mathcal{Y}^{K+1}=0)=p_{k_1k_2...k_L}\\
\vdots\hspace{7.8cm} 
\end{align*}
\begin{align*}
P(\mathcal{Y}^1>0,\mathcal{Y}^2>0,...,\mathcal{Y}^{K+1}>0)=p_{12...K+1}\\
\sum_{\substack{1\leq k_1<k_2<...k_L\leq K+1 \\ 1\leq L\leq K+1}} p_{k_1k_2...k_L}=1,\hspace{1cm}
\end{align*}
where $p_{k_1,k_2,...k_L}$ is the probability of the elements ${k_1,k_2,...,k_L}$ of the vector $\mathcal{Y}_i$ being non-zero and the rest being zero. Notice that we don't consider the case of all taxa abundance being 0 because it would not be included in the analysis. In other words, a subject has to have at least one non-zero taxa to be included in the model. This is equivalent to a $(K+1)-$dimensional Bernoulli distribution conditional on at least one Bernoulli variable being 1. The discrete part of this distribution is essentially the same as the discrete part of the multivariate zero-inflated logistic-normal (MZILN) distribution described in  \cite{MZILN}. Notice that a distribution without any zero values can be treated as a special case of a zero-inflated distribution with $p_{12...K+1}=1$ and all other $p_{k_1k_2...k_L}$ being 0 in the discrete part. With that feature, this model can be also directly applied to data sets where zero-valued data points are imputed by a Pseudocount or another positive number.

Now we define the continuous part of the two-part distribution. Conditional on a subgroup of taxa being nonzero and the rest being zero as defined in the discrete part, the joint pdf function of the continuous part is defined as:
\begin{equation*}
  f(y)=\begin{cases}
    p_1g_1\big(\log(y^1)\big), & y=(\log(y^1),0,...,0)^T.\\
   \hspace{0.5cm} \vdots & \hspace{1cm} \vdots\\
    p_{k_1...k_L}g_{k_1...k_L}\big(\log(y^{k_1}),...,\log(y^{k_L})\big), & \hspace{-0.2cm}y=(0,.,\log(y^{k_1}),0,.,0,\log(y^{k_L}),.,0)^T.\\   
   \hspace{0.5cm} \vdots & \hspace{1cm} \vdots\\
   p_{1...K+1}g_{1...K+1}\big(\log(y^1),..,\log(y^{K+1})\big), & y=(\log(y^1),..,\log(y^{K+1}))^T,
  \end{cases}
\end{equation*}
where $g_{k_1...k_L}(\cdot)$ is the density function of a $L$-dimensional multivariate normal distribution with mean $\mathcal{A}\mu$ and variance matrix $\mathcal{A}\Sigma\mathcal{A}^T$. Here $\mathcal{A}$ is a $L\times (K+1)$ matrix with the $l$th row,  $l=1,...,L$, equal to the $ k_l$th row of the $(K+1)\times (K+1)$ identity matrix, $\mu=(\mu^1,...,\mu^{K+1})^T$ is an unknown $(K+1)-$vector of means and $\Sigma$ is a $(K+1)\times (K+1)$ variance matrix. In other words, the mean vector of $\big(\log(\mathcal{Y}^{k_1}),...,\log(\mathcal{Y}^{k_L})\big)$ is $(\mu^{k_1},...,\mu^{k_L})$ which is the subvector of $\mu$ indexed by $(k_1,k_2,...,k_L)$ and its variance matrix is the submatrix of $\Sigma$ with the rows and columns indexed by $(k_1,k_2,...,k_L)$. The density function $f(y)$ includes the term $p_{k_1k_2...k_L}$ because it is essentially a density function conditional on  $\big(\mathcal{Y}^{k_1},...,\mathcal{Y}^{k_L}\big)$ being non-zero the rest of all taxa being zero. With the above definition of discrete and continuous parts, we complete the describtion of the two-part distribution which involves quite a lot of parameters including the mean vector $\mu$, the variance matrix $\Sigma$ and the discrete probability mass parameters  $p_{k_1k_2...k_L},1\leq k_1<k_2<...<k_L\leq K+1, 1\leq L\leq K+1$. The number of $p_{k_1k_2...k_L}$'s could be as many as $2^{K+1}-2$ because it needs to cover all possible scenarios of any subset of $\mathcal{Y}_i$ being non-zero. Unlike the MZILN distribution for a compositional vector in the standard simplex space $s^K$ \citep{MZILN}, the vector $\mathcal{Y}_i$ here is not in the simplex space.

\subsection{Parameters of interest}
Oftentimes, the goal of a study is to investigate the associations of microbiome abundance with other covariates such as the environmental exposures that could change microbiome composition. Suppose there are $Q$ covariates of interest, denoted by the $Q$-dimensional vector $X_i$, for the association test. Our approach can allow a large number of covariates in the model (i.e., $Q>N$). Suppose there are other $S$ covariates (e.g., confounders) that will also be included in the model, but their associations with microbiome will not be examined. The number of confounders can also be large (e.g., $S>N$). These potential confounders are denoted by $W_i$, a $S$-dimensional vector. In this paper, we are focusing on the association between $X_i$ and the microbial abundance conditional on presence. Based on the previous two-part distribution, we use the following equations to model the association:
\begin{equation}\label{eq:1b}
\log(\mathcal{Y}_i^k)|\mathcal{Y}_i^k>0=\beta^{0k}+X_i^T\beta^k+W_i^T\gamma^k+Z_i^Tb_i+\epsilon_i^k,\hspace{0.2cm}k=1,...,K+1,
\end{equation}
where the vertical line "$\vert$" means "conditional on" since the natural-log function $\log(\cdot)$ can not be applied to 0 (which will be suppressed herein for simplicity), $b_i$  are the random effects that can address the heterogeneity (e.g., biological variation) across subjects on top of the random error $\epsilon_i^k$. Here $Z_i$ is the design matrix for random effects $b_i$ which has a normal distribution with mean $\mathbf{0}$ and its variance matrix does not have to be specified. In a later section we will see that the assumption of normal distribution for $b_i$ can be relaxed. The vector (or matrix) $\mathbf{0}$ denotes a vector (or matrix) of $0$'s with appropriate dimension(s). Let $\sigma_k$ denote the standard deviation of $\epsilon_i^k$. This model can be also considered as a mixture model since the marginal distribution of $\log{(\mathcal{Y}_i^k)}$ is a linear mixture of normal distributions (of $\epsilon_i^k$) over another normal distribution (of $b_i$). Conditional on presence of taxon $k$, the parameter vector $\beta^k$ quantifies average change in the abundance of taxon $k$ on log scale given one unit change in covariates contained in $X_i$. Notice that model (\ref{eq:1b}) implies that 
\begin{align*}
&\mu_i^k=\beta^{0k}+X_i^T\beta^k+W_i^T\gamma^k,\hspace{0.1cm}i=1,\dots,N;\hspace{0.1cm}k=1,\dots,K+1,\\
&\Sigma_i=\text{diag}(\sigma_1^2,\dots,\sigma_{K+1}^2)+{\bf 1}_{K+1}Z_i^TVar(b_i)Z_i {\bf 1}_{K+1}^T,\hspace{0.1cm}i=1,\dots,N,
\end{align*}
where $\mu_i^k$ and $\Sigma_i$ were defined in Section \ref{ss:ZILoN}, $\text{diag}(\sigma_1^2,\dots,\sigma_{K+1}^2)$ is the diagonal matrix with $\sigma_1^2,\dots,\sigma_{K+1}^2$ being the diagonal elements, $Var(b_i)$ is the variance matrix of the random effect $b_i$ and ${\bf 1}_{K+1}$ is the $(K+1)-$dimensional vector of one's.

\section{Parameter estimation}\label{s:estimation}
Our target of inference is $\beta^k,k=1,\dots, K+1$. In real studies, the true taxa abundances in an ecosystem (e.g., gut), denoted by $\mathcal{Y}_i$ previously, usually cannot be observed because only a small portion of the ecosystem (e.g., stool sample) is used to produce the sequencing data. So what can be observed for the $i$th subject and $k$th taxon is $Y_i^k=C_i\mathcal{Y}^k_i$ where $C_i$ is the unknown proportion and takes value between 0 and 1. It is straightforward to see that $C_i$ is directly related to sequencing depth (i.e., library size). Let $Y_i=(Y_i^1,...,Y_i^{K+1})^T$ denote the observed vector. The unknown variable $C_i$ could cause at least two challenges, the first of which is its confounding effect \citep{Weiss2017}. This can be seen by plugging the observed abundance $Y_i^k$ into equation (\ref{eq:1b}) and the resulted equation becomes:
\begin{equation*}
\log(Y_i^k)=\log(C_i)+\log(\mathcal{Y}_i^k)=\log(C_i)+\beta^{0k}+X_i^T\beta^k+W_i^T\gamma^k+Z_i^Tb_i+\epsilon_i^k,\hspace{0.2cm}k=1,...,K+1,
\end{equation*} 
where (log-transformed) $C_i$, as a covariate in the regression equation, could be a confounder for the association of (log-transformed) $Y_i^k$ with $X_i$ when $C_i$ is associated with $X_i$ which would be true if sequencing depth is associated with $X_i$. Without appropriately accounting for the effect of $C_i$, the estimate of $\beta^k$ could be distorted toward overestimation which leads to high false positive rate or underestimation which leads to high false negative rate. The second challenge due to $C_i$ is data dispersion. It could be overdispersion or underdispersion depending on the distribution of $C_i$. For example, in the case that $\mathcal{Y}^k_i$ and $C_i$ are independent (or weakly dependent), it is straightforward to show (See Appendix for proof) that
\begin{align}\label{eq:dispersion}
\text{var}(C_i)\Big(\text{var}(\mathcal{Y}^k_i)+\big(E(\mathcal{Y}^k_i)\big)^2\Big)\leq \text{var}(Y^k_i)\leq E(C^2_i)\Big(\text{var}(\mathcal{Y}^k_i)+\big(E(\mathcal{Y}^k_i)\big)^2\Big).
\end{align}
Overdispersion happens because of the left-hand side of the inequality. For example, $\text{var}(Y^k_i)$ will be larger than $\text{var}(\mathcal{Y}^k_i)$ when $E(\mathcal{Y}^k_i)\ge \text{var}(\mathcal{Y}^k_i)$ and $\text{var}(C_i)E(\mathcal{Y}^k_i)>1$. This could explain the enormous variation of total sequencing reads across subjects commonly observed in real studies. From the right-hand side of the above inequality, we can see that $\text{var}(Y^k_i)$ could be much smaller than $\text{var}(\mathcal{Y}^k_i)$ when $E(C^2_i)$ is very small and severe underdispersion could happen in such cases. For example, when $E(C^2_i)$ is extremely small which implies that the value of $C_i$ is likely to be extremely small, $Y_i^k$ will take value close to zero and it will be difficult to observe positive abundance of $Y_i^k$ which can explain why there are so many 0's in real datasets and some taxa have nearly zero dispersion. 

\subsection{Known reference taxon}
We propose a novel method that can handle both confounding and data dispersion issues caused by unknown $C_i$. This approach involves identifying an optimal reference taxon whose log-transformation is (conditionally) independent of the covariates of interest conditional on the presence of the taxon. For illustration, let's first assume that we know there is such a taxon independent of $X_i$ and it is set to be the reference taxon. Without loss of generality, we label this reference taxon as $K+1$. We will explain the case with unknown reference taxon later. By taking the log-ratio of a taxon, say taxon $k$, over the reference taxon, we have:
\begin{align*}
\log(Y_i^k/Y_i^{K+1})&=\log(Y_i^k)-\log(Y_i^{K+1}) \\
&=\log(C_i)+\log(\mathcal{Y}_i^k)-\log(C_i)-\log(\mathcal{Y}_i^{K+1} )\\
&=\beta^{0k}-\beta^{0,K+1}+X_i^T(\beta^k-\beta^{K+1})+W_i^T(\gamma^k-\gamma^{K+1})+\epsilon_i^k-\epsilon_i^{K+1},
\end{align*}
where $\log(C_i)$ is canceled out, and thus the impact of the unobserved $C_i$ is limited in our model. Notice that $Z_i^Tb_i$ is also canceled out and thus the distribution of $b_i$ does not have to be specified and it can have a non-normal distribution. Because the (log) reference taxon is independent of $X_i$, we have $\beta^{K+1}=\mathbf{0}$. The above equation becomes:
\begin{equation}\label{eq:lr}
\log(Y_i^k/Y_i^{K+1})=\beta^{0k}-\beta^{0,K+1}+X_i^T\beta^k+W_i^T(\gamma^k-\gamma^{K+1})+\epsilon_i^k-\epsilon_i^{K+1}.
\end{equation}
From equation (\ref{eq:lr}), we can see that the log-ratio transformed data can be used to estimate the re-parameterized parameter vector $((\beta^{0k}-\beta^{0,K+1})^T,(\beta^k)^T,(\gamma^k-\gamma^{K+1})^T)^T$ from which the estimate of $\beta^k$ can be extracted. Equation (\ref{eq:lr}) also shows that $\log(Y_i^k/Y_i^{K+1})$ follows a normal distribution conditional on both $Y_i^k$ and $Y_i^{K+1}$ being non-zero because the two error terms $\epsilon_i^k$ and $\epsilon_i^{K+1}$ are independent and have normal distributions. Actually the vector $\big(\log(Y_i^1/Y_i^{K+1}), \log(Y_i^2/Y_i^{K+1}),...,\linebreak\log(Y_i^K/Y_i^{K+1})\big)$ follows a multivariate normal distribution conditional on all $Y_i^k$'s, $k=1,...,K+1$, being non-zero. Notice that 
\begin{align*}
\log(Y_i^k/Y_i^{K+1})=\log\bigg(\frac{Y_i^k}{\sum_{j=1}^{K+1}Y_i^j}\bigg/\frac{Y_i^{K+1}}{\sum_{j=1}^{K+1}Y_i^j}\bigg),k=1,...,K
\end{align*} 
The right-hand side of the above equation is actually the ratio of the two compositional proportions for taxa $k$ and $K+1$. Taken together, the composition vector $\bigg(\frac{Y_i^k}{\sum_{j=1}^{K+1}Y_i^j},...,\frac{Y_i^{K+1}}{\sum_{j=1}^{K+1}Y_i^j}\bigg)$  follows a multivariate zero-inflated logistic normal (MZILN) distribution as described in \cite{MZILN}. Therefore, the parameter vectors $\beta^k,k=1,...,K$ can be estimated with the approach proposed in \cite{MZILN} where standard regularization approaches such as LASSO \citep{Tibshirani2011a}, MCP \citep{Zhang2010a} and SCAD \citep{Fan2001} for association selection, and high-dimensional inference approaches (\citealp{Javanmard2014};\citealp{Zhang2014};\citealp{Cai2017};\citealp{HDCI}) can be incorporated to provide valid point estimates and confidence intervals for the parameters. 

\subsection{Unknown reference taxon}
In practice, we do not know which taxa are independent of which covariates. We will refer to those taxa independent of all covariates contained in $X_i$ as independent taxa and those taxa associated with any covariate in $X_i$ as associated taxa hereafter. We assume there are at least two independent taxa among all the taxa of interest. Later we will see that the independent taxon is not identifiable if there is only one such taxon. If we are able to identify an independent taxon, we can proceed with estimating the parameters as described in the previous section, and thus the task becomes to find an independent taxon that can be as the reference taxon. Taxa can be divided into two sets based on the association with $X_i$: we call the set of associated taxa (with any covariate in $X_i$) set A, and the set of independent taxa set B. It is unknown which taxon belongs to which set. So there are two possible scenarios for randomly selecting a reference taxon: it is either from set A or set B. It is obvious that $\beta^k=\mathbf{0}$ for taxa in set B, and thus the log-ratio of any two taxa in set B is independent of $X_i$. We also know that the log-ratio of any two taxa in set A is not independent of $X_i$ and the log-ratio between a taxon in set A and a taxon in set B is not independent of the covariates. So in an ideal setting with no noise, if the reference taxon is from set B for implementing the MZILN \citep{MZILN} approach with MCP, then all taxa in set B should not be selected for the association (with any covariate in $X_i$) and all taxa in set A should be selected. On the other hand, if the reference taxon is from set A, then all taxa in sets A and B should be selected for the association. Let $m_A$ and $m_B$ denote the set sizes (number of taxa) for the two sets respectively. The set sizes $m_A$ and $m_B$ are unknown, but we know that  $m_A+m_B=K+1$ since there are $K+1$ taxa in total. If we were to run the MZILN approach with MCP $K+1$ times and each time we use a different taxon as the reference taxon, then each taxa in set B should be selected $m_A$ times for the association and each taxon in set A should be selected $K$ times. If $m_A$ and $K$ are very different, i.e., the difference $K-m_A=m_B-1$ is big, we can differentiate set A and set B by simply counting the times of each taxon being selected for the association with $X_i$. The approach will not be able to differentiate sets A and B if $m_B=1$ in which case $K-m_A=0$. This is why we need to assume there are at least two independent taxa. It is straightforward to see that the bigger $m_B$, the better for our approach. If cycling through all the taxa for choosing the reference taxon, it will be very time consuming to run the MZILN approach $K+1$ times since $K$ could be very large. A more effective approach is to randomly pick $R$ different reference taxa, say $R=40$, and then run the MZILN approach with each of the picked taxa as reference taxon. This way the MZILN is implemented only $R$ times. Each taxon in set B is expected to be selected $Rm_A/(K+1)$ times for the association and each taxon in set A is expected to be selected $k_A$ times which can be calculated as follows:
\begin{align*}
k_A=\frac{{K\choose R-1}}{{K+1\choose R}}(R-1)+\frac{{K\choose R}}{{K+1\choose R}}R=\frac{R}{K+1}(R-1)+\frac{K-R+1}{K+1}R=\frac{KR}{K+1}
\end{align*}
where $\cdot\choose \cdot$ is the binomial coefficient function, and ${K\choose R-1}/{K+1\choose R}$ and ${K\choose R}/{K+1\choose R}$ are the probabilities of each taxon in set A being chosen as one of reference taxa and not chosen as one of reference taxa respectively. The mean difference of selection times will be $k_A-\frac{Rm_A}{K+1}=\frac{(m_B-1)R}{K+1}$. So $R$ should be chosen big enough for the difference $\frac{(m_B-1)R}{K+1}$ to be detectable. For example, if it is expected that about half of the taxa should be independent of $X_i$ (i.e., $(m_B-1)\approx (K+1)/2$), then choosing $R=40$ will give a mean difference approximately of $1/2\times 40=20$ which could be big enough to differentiate sets A and B. However, it might be challenging to choose $R$ without knowing the true value of $m_B$ which could lead to unacceptable misclassification of set A. We propose to use a permutation test to control the family-wise error rate (FWER) which automatically controls false discovery rate (FDR) because FWER is always larger than or equal to FDR. More details are provided in the following algorithm to select taxa in association with $X_i$.
\begin{center}
\captionof{algorithm}{Association identification and parameter estimation}\label{alg:overall}
\begin{algorithmic}[1]
\Require Family wise error rate (FWER) for taxa selection $\alpha$, number of randomly picked reference taxa $R$, number of permutations $P$
\Algphase{Phase 1a - Association identification}
\Ensure To obtain the count of each taxon being selected for the association with $X_i$.
\State\label{alg:taxa1}
Randomly pick $R$ taxa as the reference taxa set. These taxa may or may not be associated with $X_i$. Let $(Y^{T_1},\dots,Y^{T_R}), 1\le T_1<T_2<\dots<T_R\le K+1$ denote these taxa.
\State\label{alg:taxa2}
Set $r=1$ and the initial $(K+1)-$dimensional count vector $Z=(0,...,0)$ where all elements are zero. 
\State\label{alg:taxa3}
Set $Y^{T_r}$ as the reference taxon and implement the MZILN approach with MCP using the selected reference taxon. This gives sparse estimates of the parameters of interest: $\hat\beta^1,\dots, \hat\beta^{K+1}$ where $\hat\beta^{T_r}=\mathbf{0}$ because $Y^{T_r}$ is the reference taxon.
\State\label{alg:taxa4}
Record the taxa selection with the vector $Z_r$ given by $Z_r=\bigg(1_{(\hat\beta^1\neq\mathbf{0})},...,1_{(\hat\beta^{K+1}\neq\mathbf{0})}\bigg)$ where $\beta^k$'s are vectors when $Q>1$ and scalars when $Q=1$ which corresponds to the case when only one covariate is of interest for the association examination.
\State\label{alg:taxa5}
Set $r=r+1$ and $Z=Z+Z_r$, and then repeat steps \ref{alg:taxa3} and \ref{alg:taxa4} until $r$ reaches $R$ (e.g., $R=40$). The  vector $Z$ contains the count of each taxon being selected for the association with $X_i$.
\Algphase{Phase 1b - Association identification}
\Ensure Permutation to find a threshold to divide the counts in $Z$ in order to identify set A.
\State\label{alg:permu1}
Set $p=1$.
\State\label{alg:permu2}
Randomly permute the rows of the matrix consisting of only the $X$ covariates in the data set. 
\State\label{alg:permu3}
Repeat steps \ref{alg:taxa2}-\ref{alg:taxa5} by using the same reference taxa set selected previously in Phase 1a, and then find the maximum value of the vector $Z$ and denote it by $C_p^m$.
\State
Set $p=p+1$ and repeat the above steps \ref{alg:permu2} and \ref{alg:permu3} until $p$ reaches $P$ (e.g., $P=40$). And then find the $100(1-\alpha)$th percentile of the vector $(C_1^m,...,C_P^m)$ and denote it by $C^\alpha$ which is the threshold to differentiate sets A and B.
\State\label{alg:permu4}
Those taxa with counts in the vector $Z$ larger than or equal to $C^\alpha$ belong to set A and the others will be considered to belong to set B. 
\Algphase{Phase 2 - Parameter estimation}
\State\label{alg:param1}
Pick an independent taxon in set B as the final reference taxon, for example, a taxon with the smallest count in vector $Z$. One can also establish some criteria (see Section \ref{ss:criteria} in the Appendix for example) to choose a good independent reference taxon.
\State\label{alg:param2}
With the chosen reference taxon from step \ref{alg:param1}, implement MZILN along with a high-dimensional inference approach \citep{HDCI} to obtain the final estimates and confidence intervals (CI) for $\beta^1,\dots,\beta^{K+1}$. 
\end{algorithmic}
\end{center}

\section{Simulation}\label{s:simu}
\subsection{Association identification}
Extensive simulations were carried out to assess the performance of our approach in comparison with five established existing approaches: ANCOM \citep{ancom}, DESeq2 \citep{deseq2}, edgeR \citep{edge}, Wilcoxon rank sum test and ZIG \citep{zig} where DESeq2 and edgeR are popular approaches for analyzing RNA-seq data and they can be generalized to analyze microbiome data \citep{holmes,Weiss2017}. To demonstrate the robustness of our approach with respect to mis-specification of our model (\ref{eq:1b}), the simulation data was generated under the same setting as in the paper that proposed the ANCOM approach \citep{ancom}. The only change we made is that the variables $C_i$ become associated with the group assignment such that $C_i$ is a confounder of the association of interest. In our simulation, 100 data sets were generated. In each data set, there are 50 subjects divided into two groups with each group having approximately 25 subjects. This corresponds to a univariate covariate variable $X$ (ie, $Q=1$) following a Bernoulli distribution with the probability parameter being 0.5. $W$ is empty since there are no other covariates except the group variable in the model. $500$ taxa were generated in each data set and $25\%$ are assumed to have different mean abundances across the two groups. The true taxa abundance of each taxon in group 1 was generated using a Poisson distribution with the Poisson mean parameter $\lambda_j,j=1,...,500,$ generated from a gamma distribution $\Gamma (a,1)$. The parameter $a$ has three possible values: $50$, $200$ and $10000$ to represent low, medium and high abundance taxa. To mimic a real data scenario, the data was generated such that $10\%$ of the taxa had high abundance, $30\%$ medium abundance, and $60\%$ low abundance. For group 2, those taxa that have the same mean abundance as group 1 were generated with the same distribution as in group 1. Those taxa that have different means than group 1 were generated with Poisson distributions having means equal to $\lambda_j+\lambda_j^*$ where $\lambda_j^*$ was the difference of mean between group 1 and 2 and generated from a uniform distribution over the interval $(u_1, u_2)$ which is chosen to be $(100, 150)$, $(200, 400)$ or $(10000, 15000)$ to represent low, medium and high difference respectively. Among those taxa that have different means between the two groups, $60\%$, $30\%$ and $10\%$ were set to have low, medium and high differences respectively. The parameter values for $(\lambda_j,\lambda_j^*),j=1,...,500,$ were fixed for the data generation across the 100 data sets.

After the true taxa abundance $\mathcal{Y}_i^k, k=1,...,K+1$ was generated for each subject $i$ as described above, we still need to generate $C_i$ to obtain the observed abundance, $Y^k_i=[C_i\mathcal{Y}_i^k]$ where $[\cdot]$ means extracting the integer part of the number. The variable $C_i$ is allowed to be associated with the group variable which is the only difference between our setting and the setting in the ANCOM paper \citep{ancom} where $C_i$ has the same distribution across the two groups. We set $C_i$ to be a constant value within each group for simplicity. Let $C^1$ and $C^2$ denote its values in groups 1 and 2 respectively. Five scenarios were considered: $\text{Scenario 1:}\hspace{0.2cm}(C^1=1/30,C^2=1/30)$, $\text{Scenario 2:}\hspace{0.2cm}(C^1=1/30,C^2=1/90)$, $\text{Scenario 3:}\hspace{0.2cm}(C^1=1/18,C^2=1/90)$, $\text{Scenario 4:}\hspace{0.2cm}(C^1=1/9,C^2=1/90)$ and $\text{Scenario 5:}\hspace{0.2cm}(C^1=1/6,C^2=1/90)$. We use the ratio $C^1/C^2$ as a measure of the association between $C_i$ and the group variable $X$ and it is equal to 1, 3, 5, 10 and 15 for the five scenarios respectively. This ratio would be equal to the ratio of average library size if there are no difference in terms of total abundance between the two groups. So these ratios can cover a wide range of scenarios including very uneven (10X) library sizes between groups that have been studied in the literature \citep{Weiss2017}. Notice that the strength of the association increases from Scenario 1 to 5 where Scenario 1 corresponds to no association (i.e., no confounding) and Scenario 5 has the strongest association (i.e., strongest confounding). We studied the performance of our approach and others under the five scenarios. Four indices were used to evaluate the performance: Recall, Precision, F1 and Type I error rate (Type1) that were calculated as follows:
\begin{align*}
\text{Recall}=\frac{TP}{TP+FN}, \hspace{0.5cm}\text{Precision}=\frac{TP}{TP+FP},\hspace{0.5cm} \text{F1}=\frac{2}{\frac{1}{\text{recall}}+\frac{1}{\text{precision}}},
\hspace{0.5cm}\text{Type1}=\frac{FP}{FP+TN}
\end{align*}
where $TP$, $FP$, $FN$ and $TN$ denote true positive, false positive, false negative and true negative respectively. Recall is a measure of statistical power, the higher the better. Precision has an inverse relationship with false discovery rate (FDR) which is equal to (1-Precision), and thus the higher the Precision, the lower the FDR. F1 is the Harmonic mean \citep{Hmean} of Recall and Precision that measures the overall performance in terms of Recall and Precision. The targeted FDR level is set to be 20\% for all approaches. When implementing IFAA, we choose the FWER to be $\alpha=20\%$ such that FDR$\le$20\%, the number of random reference taxa $R=40$ and the number of permutations $P=40$. For implementing the ANCOM approach, the stringent correction option was used in the ANCOM R package throughout this paper. 

\begin{figure}[H]
\centering
\includegraphics[width =0.7\textwidth,angle=0]{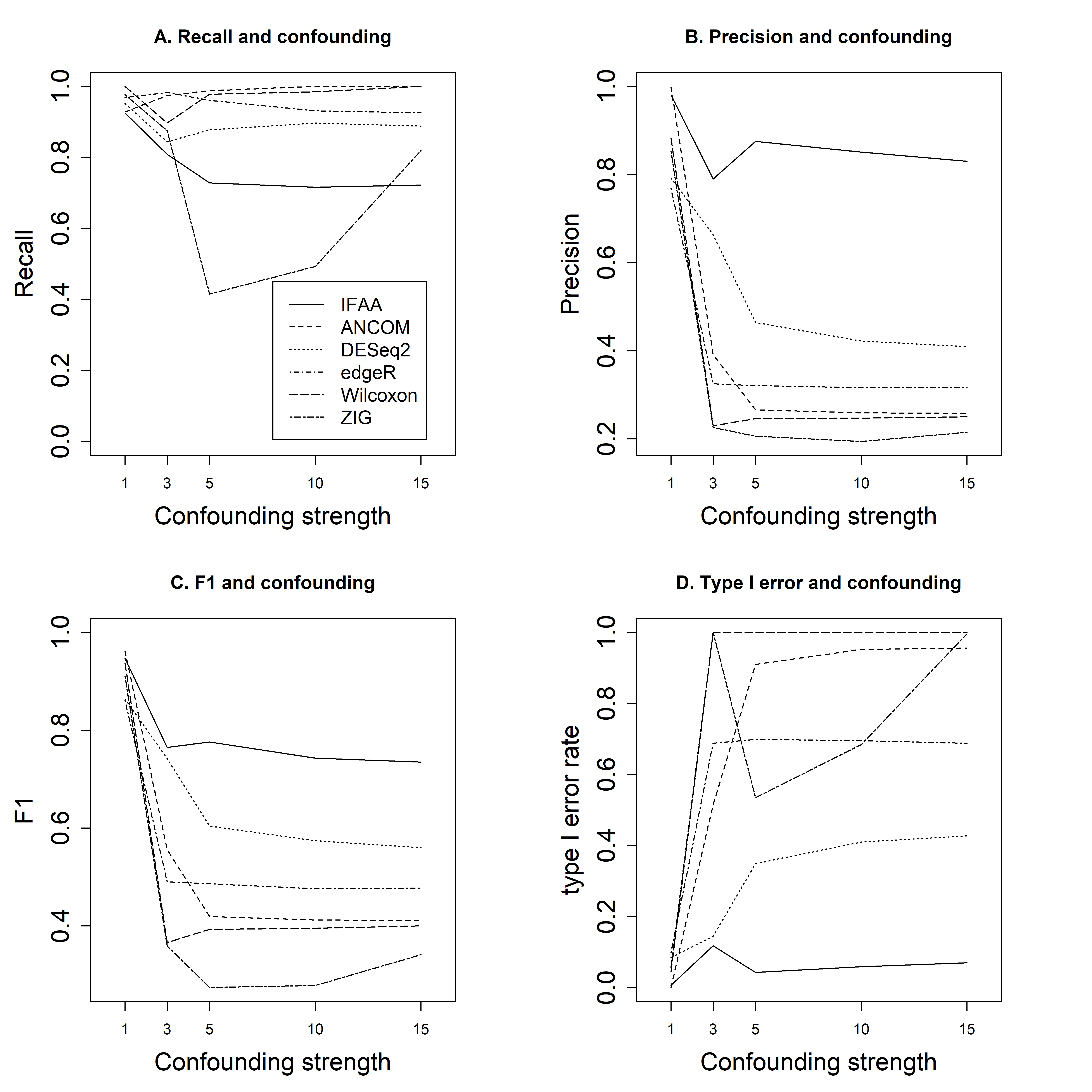}
\caption{Comparison with ANCOM, DESeq2, edgeR, Wilcoxon rank sum test and ZIG}
\label{simuComparison}
\end{figure}

We plotted the four performance measures against confounding strength as shown in Fig.\ref{simuComparison}. When there was no confounding effect (i.e., Scenario 1), all approaches had Precision rates (Fig.\ref{simuComparison}B) above or around 80\% with DESeq2 and edgeR having the lowest Precision rates of (79.2\%, 76.8\%) that translate to FDR of (20.8\%, 23.1\%) which were a little higher than the targed FDR of 20\%. All approaches had good Recall rates ($>$92\%) and good type I error rates ($<$0.1) with ANCOM and our approach (IFAA) having the smallest type I error rates when there was no confounding. As the confounding strength increases, Precision rates (Fig.\ref{simuComparison}B) dropped dramatically for all approaches except IFAA. Although the Precision rate of IFAA dropped to 79\% at Scenario 1, it stayed higher than 80\% across all other scenarios and thus achieved the desired FDR of 20\% even for Scenario 5 that had the strongest confounding effect of $C_i$. Precision rates of all other approaches dropped to below 67\% at Scenario 2, below 47\% at Scenario 3, below 43\% at Scenario 4 and below 41\% at Scenario 5 which translates to $>$59\% FDR rate that almost tripled the desired FDR of 20\%. ZIG and Wilcoxon rank sum test had the worst performance in terms of Precision rate which dropped to below 26\% starting form Scenario 2 and that translates to FDR$>$74\%. The Recall rate of IFAA (Fig.\ref{simuComparison}A) dropped from 0.93 to 0.81 at Scenario 2 and further dropped to 0.72 and remained stable after departing from Scenario 2. F1 score, the measure of overall performance in terms of Recall and Precision, of IFAA (Fig.\ref{simuComparison}C) had the best values in the presence of confounding and outperformed all the other approaches by a big margin starting from Scenario 3. DESeq2 ranked number 2 in terms of F1 score in the presence of confounding effects.  As the confounding strength increases, ZIG had the worst F1 score because of its lowest Precision rate and big drops of Recall rate at Scenarios 3 and 4. ZIG showed a strange behavior of Recall rate. Its Recall rate dropped to 42\% at Scenatio 3 and then bounced back to 82\% at Scenario 5. We also examined the type I error rate in relation with the confounding strength (Fig.\ref{simuComparison}D). IFAA had the lowest type I error rate ($<$0.12) for all scenarios with confounding effects. All other approaches had highly inflated type I error rates as the confounding strength increases. Some even had type I error rate inflated to above 0.95 at Scenario 5 such as Wilcoxon rank sum test, ZIG and ANCOM. Type I error rates of DESeq2 and edgeR were inflated to 0.43 and 0.69 respectively at Scenario 5. 

\subsection{Parameter estimation}
Once sets A and B were identified, we chose a taxon from set B that had the smallest count in vector $Z$ as the final reference taxon to obtain parameter estimates in Phase 2 of the Algorithm. As far as we know, there is no existing approach that can provide association parameter estimates regarding AA, so we did not have any existing approaches to compare with. We checked the estimation bias of IFAA for those truly non-zero values of $\beta^k,k=1,\dots,K+1$ (see table \ref{Tab:EP}). The true parameter value for $\beta^k$ was calculated as the $E\big(\log(\mathcal{Y}^k)|X=1,\mathcal{Y}^k>0\big)-E\big(\log(\mathcal{Y}^k)|X=0,\mathcal{Y}^k>0\big)$. Results showed that the mean magnitude of all biases stayed fairly stable across all scenarios including the case with strongest confounding effect. The estimates were expected to be biased because model (\ref{eq:1b}) was severely mis-specified in the data generation. This performance was not too bad given that the results were fairly robust with respect to different confounding effects.

\begin{table}[H]
\caption{Estimation performance}\label{Tab:EP}
\begin{center}
 \begin{tabular}{|c|c|c|c|c|c|} 
 \hline
 Confounding strength&mean of true parameter values & mean magnitude of biases & Bias\% \\
 \hline
 1 & 1.74 & 0.21 & 11.95 \\
 \hline
 3 & 1.74 & 0.24 & 14.06 \\
 \hline
 5 & 1.74 & 0.18 & 10.57 \\
 \hline
 10 & 1.74 & 0.18 & 10.11\\
 \hline
 15 & 1.74 & 0.20 & 11.55\\
 \hline
\end{tabular}
\end{center}
\end{table}

\section{Real study applications}\label{s:app}
\subsection{New Hampshire Birth Cohort Study (NHBCS)}
\begin{figure}[H]
  \begin{center}
  \includegraphics[width=0.9\textwidth,angle=0]{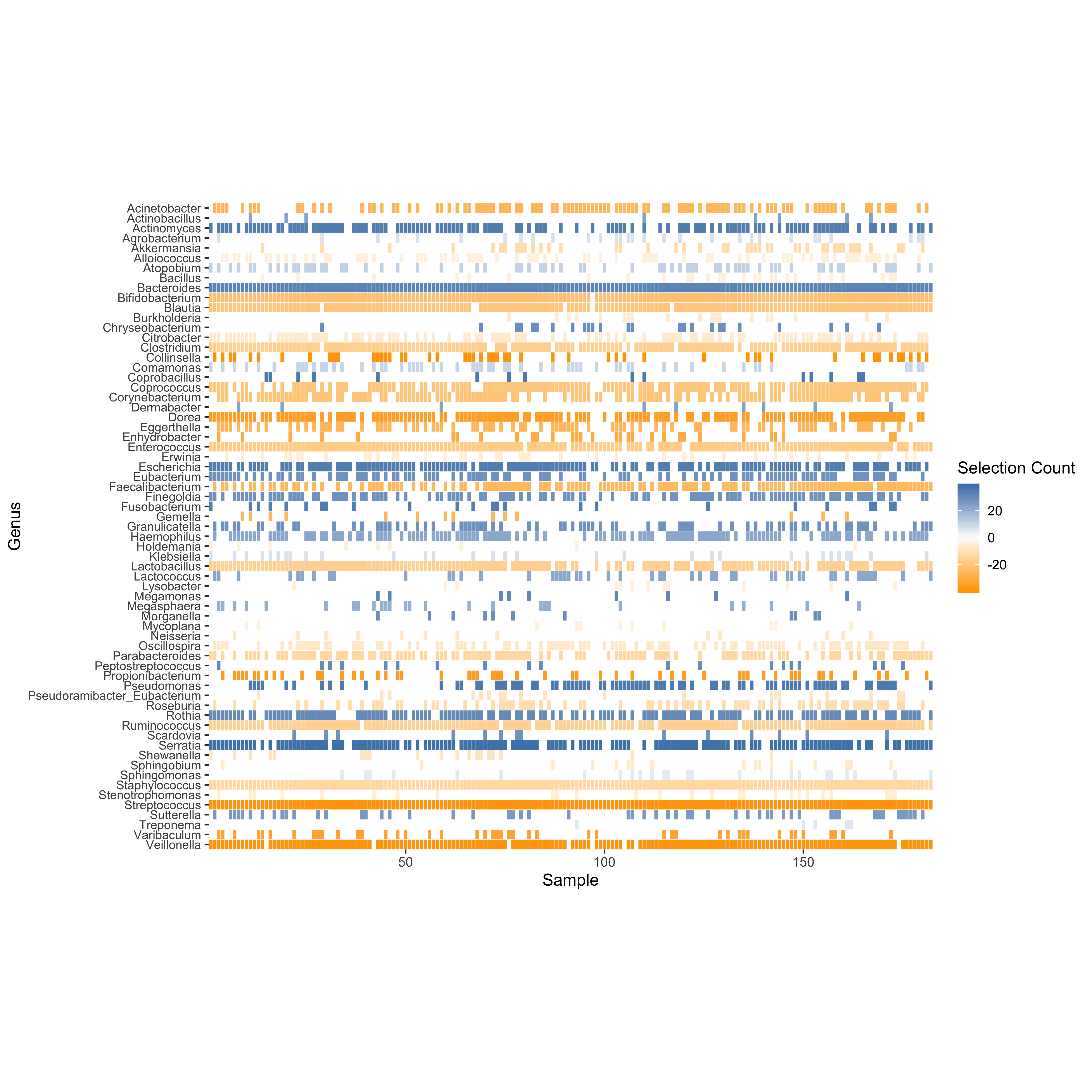}
  \end{center}
  \caption{Assocation heatmap for the NHBCS study. Blue and orange denote positive and negative associations with arsenic level respectively. Selection count from Phase 1 of the algorithm determine the darkness of the colors for all genera. Negative sign means negative association. Absence is coded as 0. Genera selected less than 5 times are not included. Genera are labeled on the vertical axis and samples are labeled on the horizontal axis.}
  \label{heatNHBCS}
\end{figure}

The NHBCS is a large NIH-funded ongoing longitudinal epidemiological project to study the health impacts of environmental exposures such as arsenic in mothers and their children \citep{Farzan2013}. Pregnant mothers were recruited to the study at approximately 24 to 28 weeks of gestational age and longitudinal data are collected from both mothers and babies at followed up time points. We applied our approach in the NHBCS study to examine the association between {\it in utero} arsenic exposure measured by maternal urinary arsenic concentrations \citep{Farzan2016} during pregnancy and the infant gut microbiome. In our analysis, the natural log-transformed total {\it in utero} arsenic level \citep{Nadeau2014} was the exposure variable $X$ and gut microbiome of infants at 6 weeks of age was the outcome variable. Delivery mode (vaginal VS. C-Section) and feeding type (Breast fed VS. others) were adjusted as potential confounders in the model (i.e., $W_i$ in equation (\ref{eq:1b})). The gut microbiome data was measured in DNA extracted from infant stool samples using 16S rRNA sequencing of the V4-V5 hypervariable regions (\citealp{Madan2016};  \citealp{MZILN}). Sequencing reads were quality checked and clustered into operational taxonomic units as described previously \citep{Madan2016}. After quality control and data cleaning, there were 182 subjects and 218 genera available in the data set. About 85\% of the microbiome data points were zero. AA of genera were analyzed as the outcome variables. Our model found two genera: {\it Collinsella} and {\it Serratia} that were significantly associated with {\it in-utero} arsenic concentrations. FWER was controlled at 30\%, 40 permutations were used and 40 reference taxa were randomly chosen in Algorithm \ref{alg:overall} (i.e., $\alpha=0.30$, $P=40$, $R=40$). It took about 73 minutes to finish running the analysis on a 8-core Windows 10 machine. The regression coefficients estimated from IFAA were -1.17 and 1.06 respectively meaning that one unit increase on the log-scale of {\it in-utero} arsenic exposure level would lead to 69\% reduction in the absolute abundance of {\it Collinsella} and 1.9-fold increase in the absolute abundance of {\it Serratia} on average in the entire gut conditional on presence of these genera. The 95\% CI calculated with a Bootstrap Lasso + Partial Ridge method \citep{HDCI} for the regression coefficients were (-1.42, -0.10) and (-0.18, 0.79) respectively without multiple testing correction. While {\it Collinsella} is an innovative finding, {\it Serratia} has been linked to arsenic in the literature \cite{Lukasz2014}. To give a full picture of all associations, a heatmap (Figure \ref{heatNHBCS}) was also constructed to show the number of times each genus was selected for the association with arsenic level in Phase 1 of the algorithm. These selection counts can be considered as measures of the strengths of associations. For comparison, we analyzed the data with the ANCOM method as well. Since the ANCOM R package does not allow adjusting for potential confounders, the raw associations between the arsenic variable and the gut microbiome were tested using ANCOM. It did not find any genera that are statistically significantly associated with the arsenic variable at the same FDR rate of 30\%. We also applied the nonparametric Spearman correlation for testing the raw correlations between the arsenic variable and RA and it did not identify any taxa which suggests that the signal-to-noise ratio in this dataset might be weak (which could be due to the high data sparsity with 85\% zeros) since simple nonparametric tests tend to overidentify associated taxa but it did not detect any assoicated taxa in this dataset. The Spearman correlation test for correlations between the arsenic variable and AA did not result in any significant associations either. We did not compare with DESeq2, EdgeR and ZIG in this application because they were developed for differential abundance analysis between two groups whereas the exposure variable here, {\it in-utero} arsenic level, is a continuous variable. 

\subsection{VSL\#3 mouse model}
\begin{figure}[H]
  \begin{center}
  \includegraphics[width =0.8\textwidth,angle=0]{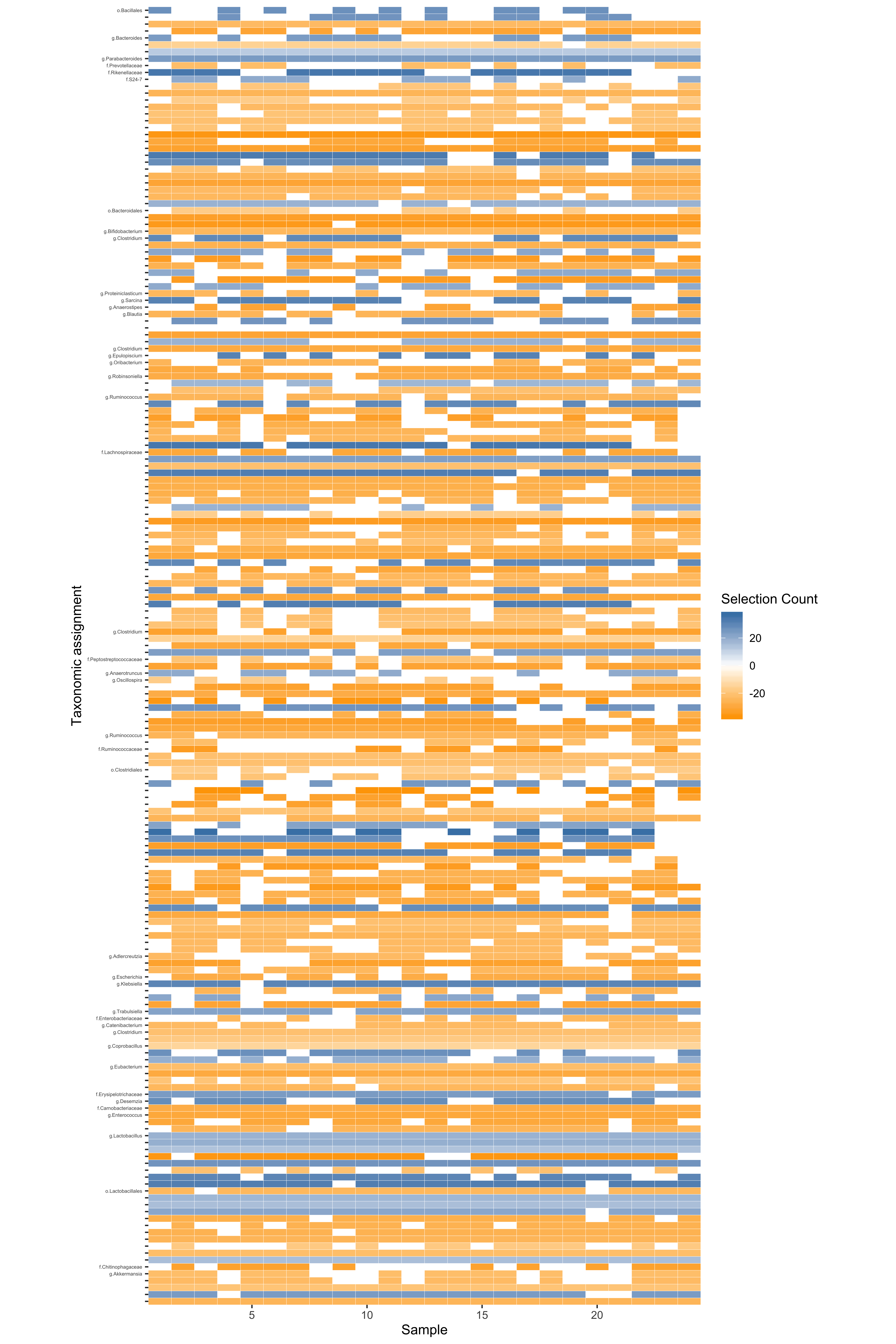}
  \end{center}
  \caption{Assocation heatmap for the VSL\#3 study. Selection count from Phase 1 of the algorithm determine the darkness of the colors for all OTUs. Negative sign means negative association. Absence is coded as 0. OTUs selected less than 15 times are not included in the figure. Taxonomic assignment is labeled on the vertical axis and the unlabeled OTUs belong to the taxon on its top. Samples are labeled on the horizontal axis.}
  \label{heatMouse}
\end{figure}

VSL\#3 is a commercially available probiotic cocktail (Sigma-Tau Pharmaceuticals, Inc.) of eight strains of lactic acid-producing bacteria. In a mouse model, Arthur et al. \citep{Arthur2013} studied the ability of VSL\#3 to alter the colonic microbiota and decrease inflammation-associated colorectal cancer when administered as interventional therapy after the onset of inflammation. In this study, there were totally 23 mice of which 10 were treated with VSL\#3 and 13 served as control. Gut microbiome data were collected from stools at the end of the study with 16S rRNA sequencing \citep{lzgMediation}. There were 362 OTUs in total in the data sets after quality control and data cleaning. About 40\% of the OTU abundance data points were zero. In this application, we are interested in the association between the gut microbiome and the dysplasia score (the higher the worse) which is a continuous variable measuring the abnormality of cell growth. AA of OTUs were analyzed as the $Y$ variable in the model. The treatment variable was adjusted as a potential confounder for this association in the analysis (i.e., $W_i$ in equation (\ref{eq:1b})). Again, FWER was controlled at 30\%, 40 permutations were used and 40 reference taxa were randomly chosen in Algorithm \ref{alg:overall} (i.e., $\alpha=0.30$, $P=40$, $R=40$). It took about 125 minutes to finish the analysis on a 8-core Windows 10 machine. Two OTUs were found to be significantly associated with the dysplasia score with one OTU assigned to the kingdom Bacteria and and the other OTU assigned to family S24-7 within the order Bacteroidales. The regression coefficients for the two OTUs were -1.18 (95\% CI: -1.04, -0.12) and -0.87 (95\% CI: -1.75, -0.78) respectively where the CI's were calculated using the Bootstrap LPR method \citep{HDCI}. The negative associations suggest that these OTUs are associated with reduced dysplasia score and, on average, one unit increase of the dysplasia score is associated with 65\% and 58\% reduction in the absolute abundance of the two OTUs in the entire gut conditional on the presence of these OTUs. These findings are consistent with associations of Bacteroidales and S24-7 with intestinal tumorigenesis reported in the literature  (\citealp{Braten2017}; \citealp{Rudi2017}). To give a full picture of all associations, a heatmap (Figure \ref{heatMouse}) was also constructed to show the number of times each OTU was selected for the association with dysplasia score in Phase 1 of the algorithm. We applied the ANCOM approach to test the raw associations between the dysplasia score and microbiome since its R package does not allow adjusting for potential confounders. ANCOM did not identify any OTUs at the same FDR rate of 30\%. The nonparametric Spearman correlation test identified 68 taxa AA which is likely to be an overidentification. When testing the correlations of RA with the dysplasia score using Spearman correlation test, 61 taxa RA were identified. Again, we did not compare with DESeq2, EdgeR and ZIG in this application because the dysplasia score a continuous variable. 

\section{Discussion}\label{s:discu}
We developed a novel approach (IFAA) that can draw inferences directly on the absolute abundance (AA) of microbial taxa in an ecosystem and provide point estimates and confidence intervals for the associations of AA with other covariates. By making inference on AA, IFAA circumvents the issues induced by the features of RA such as the CD problem and the compositional structure. IFAA can also address the possible confounding effect of sequencing depth that has been a challenging problem in the literature of microbiome research \citep{Weiss2017}. IFAA identifies microbial taxa associated with the covariates of interest (set A) and the other taxa that are not associated with the covariates (set B) with a desired false positive rate in set A in Phase 1 of the alrogithm where permutation method is used to control FDR by controlling FWER since FWER can serve as an upper bound of FDR. In Phase 2, a reference taxon from set B that is independent of the covariates is picked for the model to generated valid estimates of the associations of all taxa in set A with the covariates. When using IFAA, one does not need to impute zero sequencing reads with a Pseudocount or any other number for the analysis which can avoid bias caused by the imputation. Although imputation of zero-valued reads is not required, IFAA can still be directly applied to data sets containing imputed values for investigators who are comfortable with imputation. When there are no zeros in an imputed data set, it corresponds to the  zero-inflated log-normal distribution with $p_{12\dots K+1}=1$ and all other masses are zero in the discrete part as shown in Section \ref{ss:ZILoN}. Normalization methods such as rarefaction \citep{Weiss2017} can also be allowed in our approach to normalize the data for analysis. IFAA can also be directly applied to RA data as well because the ratio of two RA's is the same as the their AA's. This could be helpful for investigators who want to draw inference on AA with RA data. Our approach can be applied to different settings including two-group comparisons and regressions with continuous exposure variables where confounders can be adjusted in the model. IFAA can handle high-dimensional microbiome data as well as high-dimensional covariates data by incorporating regularization methods. An R package to implement IFAA can be installed directly from the github website (https://github.com/gitlzg/IFAA). 

We started with assuming normal distributions for the random errors $\epsilon_i^k$, but this assumption is not required as long as the distributions have mean of zero because the parameters are estimated using estimating equations \citep{MZILN}. This property ensures the robustness of our approach (as demonstrated in the simulation) for a broad range of distributions that could be encountered in practice under different study settings with different study populations. Although we did not study batch effect on the method in this paper, we expect it to have good performance in the presence of batch effects because the ratio of two taxa abundances does not depend on library size, and therefore it should generate robust results with respect to batch effects on library size. This is similar to the phenomenon of controlling for the confounding effect of library size as presented in the simulation study. Most existing approaches including those for RNA-seq and microarray data use a normalization procedure to deal with batch effects (\citealp{Chen2011}; \citealp{Ritchie2015}; \citealp{Gibbons2018}) and some incorporate the batch effect adjustment in the regression model for final analysis \citep{Dai2018}. We will study the performance of our approach in comparison with existing approaches in a future project.

Another implicit assumption, which is also needed in the ANCOM method \citep{ancom}, is that the observed abundance $Y^k_i$ is equal to $C_i\mathcal{Y}^k_i$ which might not be true in practice because $C_i\mathcal{Y}^k_i$ is probably not an integer most of the time. This assumption is important for using the ratios of abundances because $C_i$ can be canceled out in the ratios under this assumption. What is observed in practice is an integer, so it might make more sense to assume that $Y^k_i=[C_i\mathcal{Y}^k_i]$ where $[x]$ denotes extracting the integer part of $x$. However, it can be shown that the difference, $\log(C_i\mathcal{Y}^k_i)-\log\big([C_i\mathcal{Y}^k_i]\big)$ (given $[C_i\mathcal{Y}^k_i]\ge 1$), is bounded by $1/[C_i\mathcal{Y}^k_i]$ (see Appendix for proof), and thus the impact of this difference on the estimation of $\beta^k$'s is likely to be limited since the estimation for $\beta^k$'s is conditional on non-zero observation of the abundance. This paper focuses on studying the association of non-zero taxa with exposures. The presence/absence analysis of the microbial taxa can be treated as nuisance to the analysis of non-zero taxa \citep{MZILN} and warrants future research as a separate project. 

IFAA is flexible in terms of choosing the high-dimensional inference method in Phase 2 to obtain point estimates and confidence intervals for the parameters of interest. In this paper, we used a Bootstrap Lasso + Partial Ridge method \citep{HDCI} that requires less assumptions and can be readily applied using the R package ``HDCI'', but many other such methods can be employed in Phase 2 as well including (\citealp{Javanmard2014}; \citealp{Zhang2014}; \citealp{Cai2017}). It warrants further investigation to select an optimal high-dimensional inference approach in combination with MZILN in Phase 2 for analyzing microbiome data that have complex inter-taxa correlation structure. When there are more than one good independent reference taxa available in Phase 2 for parameter estimation, an alternative way to obtain the parameter estimates could be implementing the steps in Phase 2 for all good independent reference taxa one by one and then take the average of all estimates for the final estimates. This will likely generate more stable estimates at the cost of increased computational burden.

\section{Appendix}
\subsection{Proof for equation (3): the dispersion equation}
When $C_i$ and $\mathcal{Y}^k_i$ are independent (or weakly dependent), we prove the following equation:
\begin{align*}
\text{var}(C_i)\Big(\text{var}(\mathcal{Y}^k_i)+\big(E(\mathcal{Y}^k_i)\big)^2\Big)\leq \text{var}(Y^k_i)\leq E(C^2_i)\Big(\text{var}(\mathcal{Y}^k_i)+\big(E(\mathcal{Y}^k_i)\big)^2\Big).
\end{align*}
{\bf Proof:} We first show the inequality on the right-hand side:
\begin{align*}
&\text{var}(Y^k_i)=\text{var}(C_i\mathcal{Y}^k_i)\\
&=E(C^2_i(\mathcal{Y}^k_i)^2)-E(C_i\mathcal{Y}^k_i)^2\\
&\le E(C^2_i(\mathcal{Y}^k_i)^2)=E(C^2_i)E((\mathcal{Y}^k_i)^2)\\
&=E(C^2_i)(\text{var}(\mathcal{Y}^k_i)+E(\mathcal{Y}^k_i)^2)
\end{align*}
For the left-hand side, we have
\begin{align*}
&\text{var}(Y^k_i)=\text{var}(C_i\mathcal{Y}^k_i)\\
&=E(C^2_i(\mathcal{Y}^k_i)^2)-E(C_i)^2E(\mathcal{Y}^k_i)^2\\
&=E(C^2_i)E((\mathcal{Y}^k_i)^2)-E(C_i)^2E(\mathcal{Y}^k_i)^2\\
&=E(C^2_i)E((\mathcal{Y}^k_i)^2)-E(C^2_i)E(\mathcal{Y}^k_i)^2+E(C^2_i)E(\mathcal{Y}^k_i)^2-E(C_i)^2E(\mathcal{Y}^k_i)^2\\
&=E(C^2_i)\text{var}(\mathcal{Y}^k_i)+\text{var}(C_i)E(\mathcal{Y}^k_i)^2\\
&\ge\text{var}(C_i)\text{var}(\mathcal{Y}^k_i)+\text{var}(C_i)E(\mathcal{Y}^k_i)^2\\
&=\text{var}(C_i)\Big(\text{var}(\mathcal{Y}^k_i)+E(\mathcal{Y}^k_i)^2\Big)
\end{align*}

\subsection{Proof for the bound of the difference: $\log(C_i\mathcal{Y}^k_i)-\log\big([C_i\mathcal{Y}^k_i]\big)$}
For $[C_i\mathcal{Y}^k_i]\ge1$ which is the case we consider in the paper, let $\delta=C_i\mathcal{Y}^k_i-[C_i\mathcal{Y}^k_i]$ and thus $\delta\in[0,1)$. We have
\begin{align*}
0\le\log(C_i\mathcal{Y}^k_i)-\log\big([C_i\mathcal{Y}^k_i]\big)&=\log([C_i\mathcal{Y}^k_i]+\delta)-\log\big([C_i\mathcal{Y}^k_i]\big)\\
&=\log\bigg(1+\frac{\delta}{[C_i\mathcal{Y}^k_i]}\bigg)\\
&\le\frac{\delta}{[C_i\mathcal{Y}^k_i]}\\
&<\frac{1}{[C_i\mathcal{Y}^k_i]}.
\end{align*}
The first inequality is because $\log(1+x)\le x$ for any non-negative number $x$. So the difference could become very small when the observed absolute abundance $[C_i\mathcal{Y}^k_i]$ is large.

\subsection{Suggestive criteria for identifying the final reference taxon}\label{ss:criteria}
Since a final independent taxon is needed in Phase 2 of the algorithm to obtain parameter estimates, it might be helpful to have some criteria in place for finding a good independent taxon in set B. The following are some criteria that might be useful.
\begin{titlemize}{\textit{Suggestive criteria for identifying the final reference taxon:}}
\item The final reference taxon has $10\%$ or more non-zero abundances observed among those subjects who have two or more observed non-zero taxa.
\label{cri:1}

\item When making inference on the associations with a binary covariate, the final reference taxon has $10\%$ or more non-zero abundances observed in each group indicated by the binary covariate among those subjects who have two or more observed non-zero taxa.
\label{cri:2}

\item The final reference taxon has a small (if not zero) count contained in the vector $Z$ as calculated in step \ref{alg:taxa4} of Algorithm \ref{alg:overall}. The first tertile of the counts for all taxa in set A can be used as the threshold for good independent reference taxa. The cut at first tertile can be customized depending on the distribution of the counts in vector $Z$.
\label{cri:3}

\item The final reference taxon has enough variation for observed abundances caused by the variation of library size. For example, a taxon with sequencing reads equal to 1 in all subjects is not a good final reference taxon because its variance is 0.
\label{cri:4}
\end{titlemize}

\begin{remark}
The first two criteria are only relevant when there are zero-valued sequencing reads. If the method is applied to data sets where all zeros have been imputed by a Pseudocount or another number, these two criteria are not needed.  
\end{remark}

\begin{remark}
The reasons we only consider ``subjects who have two or more observed non-zero taxa'' are because our approach is the based on the log-ratio transformation of the taxa abundance which requires at least two non-zero taxa to calculate a ratio.
\end{remark}

\section{Funding}
This work was supported in part by US NIH grants R01GM123014, UH3OD023275, P01ES022832, P20GM104416 and U.S. EPA grant RD 83544201.

\vspace{0.1cm}
\bibliography{../../JabRef}
\bibliographystyle{apa}
\end{document}